\documentclass[a4paper]{article}

\usepackage{INTERSPEECH2020}

\usepackage{color,xcolor}
\usepackage{epsfig}
\usepackage{graphicx}

\usepackage{array}
\usepackage{booktabs}
\usepackage{colortbl}
\usepackage{multirow}
\usepackage{float}
\usepackage{caption}
\usepackage[labelformat=simple]{subcaption}

\usepackage{footnote}
\makesavenoteenv{tabular}
\makesavenoteenv{table}

\usepackage{amsmath,amsfonts,amssymb,bm}
\usepackage[super]{nth}

\usepackage{changepage}
\usepackage{extramarks}
\usepackage{lastpage}
\usepackage{setspace}
\usepackage{soul}
\usepackage{xspace}

\usepackage{url}
\usepackage{hyperref}
\hypersetup{colorlinks=True, urlcolor=black}

\usepackage{algorithm, algorithmic}
\usepackage{enumitem}
\usepackage{verbatim}
\usepackage{pifont}
\usepackage[acronyms]{glossaries}
\glsdisablehyper
\usepackage{pifont}
\newcommand{\sect}[1]{Section~\ref{sec:#1}}
\newcommand{\sectdot}[1]{Sec.~\ref{sec:#1}}

\newcommand{\eqn}[1]{Equation~(\ref{eqn:#1})}
\newcommand{\eqndot}[1]{Eqn.~(\ref{eqn:#1})}
\newcommand{\fig}[1]{Figure~\ref{fig:#1}}

\newcommand{\tbl}[1]{Table~\ref{tab:#1}}

\newcommand{\ignore}[1]{}

\newcommand{\vect}[1]{\boldsymbol{\mathbf{#1}}}

\DeclareMathOperator*{\argmax}{arg\,max}

\makeatletter
\DeclareRobustCommand\onedot{\futurelet\@let@token\@onedot}
\def\@onedot{\ifx\@let@token.\else.\null\fi\xspace}

\def\ie{\emph{i.e}\onedot}

\makeatother

\definecolor{MyDarkBlue}{rgb}{0,0.08,1}
\definecolor{MyDarkGreen}{rgb}{0.02,0.6,0.02}
\definecolor{MyDarkRed}{rgb}{0.8,0.02,0.02}
\definecolor{MyDarkOrange}{rgb}{0.40,0.2,0.02}
\definecolor{MyPurple}{RGB}{111,0,255}
\definecolor{MyRed}{rgb}{1.0,0.0,0.0}
\definecolor{MyGold}{rgb}{0.75,0.6,0.12}
\definecolor{MyDarkgray}{rgb}{0.66, 0.66, 0.66}

\title{Residual Energy-Based Models for End-to-End Speech Recognition}
\name{Qiujia Li$^{1*}$, Yu Zhang$^2$, Bo Li$^2$, Liangliang Cao$^2$, Philip C. Woodland$^1$
\thanks{$^*$Work was done while the author interned at Google.}
}

\address{
  $^1$University of Cambridge, UK, $^2$Google LLC, USA}
\email{$^1$\{ql264,pcw\}@eng.cam.ac.uk, $^2$\{ngyuzh,boboli,llcao\}@google.com}

\begin{document}

\maketitle
\begin{abstract}
End-to-end models with auto-regressive decoders have shown impressive results for automatic speech recognition (ASR). These models formulate the sequence-level probability as a product of the conditional probabilities of all individual tokens given their histories. However, the performance of locally normalised models can be sub-optimal because of factors such as exposure bias. Consequently, the model distribution differs from the underlying data distribution. In this paper, the residual energy-based model (R-EBM) is proposed to complement the auto-regressive ASR model to close the gap between the two distributions. Meanwhile, R-EBMs can also be regarded as utterance-level confidence estimators, which may benefit many downstream tasks. Experiments on a 100hr LibriSpeech dataset show that R-EBMs can reduce the word error rates (WERs) by 8.2\%/6.7\% while improving areas under precision-recall curves of confidence scores by 12.6\%/28.4\% on test-clean/test-other sets. Furthermore, on a state-of-the-art model using self-supervised learning (wav2vec 2.0), R-EBMs still significantly improves both the WER and confidence estimation performance.
\end{abstract}
\section{Introduction}
\label{sec:intro}
End-to-end trainable speech recognition models have shown promising performance in automatic speech recognition (ASR)~\cite{Chiu2018StateoftheArtSR,Tske2020SingleHA,Sainath2020ASO,Guo2020RecentDO,Li2020DevelopingRM}, especially using recurrent neural network transducers (RNN-Ts)~\cite{Graves2012SequenceTW} and attention-based encoder-decoder models~\cite{Chorowski2015AttentionBasedMF}. One common characteristic of these two types of model is that the decoder learns the conditional distribution of the current output token given all the history tokens. This leads to \emph{locally normalised} auto-regressive models where the output probability distributions are normalised per output token. The final probability of the output sequence is obtained by computing the product of a series of conditional probabilities. This allows end-to-end models to be trained efficiently using the maximum likelihood criterion and yields good ASR performance.

However, locally normalised models can in practice be sub-optimal. First, history tokens are the ground truth during training while history tokens may contain errors during inference as they are generated sequentially. This training and inference mismatch is referred to as exposure bias~\cite{Ranzato2016SequenceLT}. To this end, scheduled sampling has been proposed to allow generated tokens to appear in the history during training with a certain probability according to a specific schedule~\cite{Bengio2015ScheduledSF}. Using a larger beam during decoding or incorporating beam search heuristics can also help reduce search errors~\cite{Chorowski2017TowardsBD}. Secondly, the locally normalised model trained with maximum likelihood may not be optimal in terms of the final evaluation criterion, \ie word error rate (WER) for ASR. Sequence-level training criteria that directly minimise the number of word errors have been proposed to address this mismatch~\cite{Prabhavalkar2018MinimumWE}. Although all of the above techniques can improve model performance to some extent, the model distribution is still based on the product of locally normalised distributions, which may differ from the data distribution~\cite{Deng2020Residual}.

\emph{Residual energy-based models} (R-EBMs), which were first used for text generation~\cite{Deng2020Residual}, use energy-based models (EBMs)~\cite{LeCun2006ATO} to learn from the residual errors of an auto-regressive generator to reduce the gap between the model and data distributions. An R-EBM can also be viewed as a discriminator between generated samples from an auto-regressive model and real data samples~\cite{Deng2020Residual}. For end-to-end ASR, the R-EBM, conditioned on acoustic features, aims to distinguish the model-generated hypotheses from the ground truth transcriptions. R-EBMs are trained using conditional noise contrastive estimation (NCE)~\cite{Ma2018NoiseCE,Gutmann2010NoisecontrastiveEA}. For a given utterance, the positive sample is the ground truth whereas the n-best hypotheses are taken as the negative samples. During inference, the auto-regressive model first generates a list of hypotheses for each utterance, and the best hypothesis from the joint model can be obtained by using the combined score of the log-likelihood from the auto-regressive model and the negative energy value from the R-EBM.

As a discriminator between correct and erroneous hypotheses, R-EBMs can also produce \emph{utterance-level confidence scores} for end-to-end ASR models. Some downstream tasks only require utterance-level confidence scores, such as data selection for semi-supervised learning~\cite{Park2020ImprovedNS,Zhang2020PushingTL}, and hypothesis-level model combination~\cite{Qiu2021LearningWC}. Compared to token or word level confidence, direct modelling of utterance-level confidence scores implicitly takes deletion errors into account, and does not require calibration of the confidence scores (e.g. using piece-wise linear mapping~\cite{Evermann2000LargeVD}) before taking the average for utterance-level scores. Previously, many model-based methods have been used for confidence estimation for both conventional~\cite{Kalgaonkar2015EstimatingCS,DelAgua2018SpeakerAdaptedCM,Li2019BidirectionalLR,Ragni2018ConfidenceEA} and end-to-end ASR~\cite{li2020confidence,Qiu2021LearningWC,Oneata2021AnEO,Kumar2020UtteranceCM,Woodward2020ConfidenceMI,Qiu2021MultiTaskLF}, and $n$-best re-ranking models~\cite{Ogawa2018RescoringNS,Li2019IntegratingSA,Sainath2019TwoPassES,Variani2020NeuralOS} have been proposed to improve WER. The R-EBM is a single model that can improve speech recognition performance and utterance-level confidence estimation performance at the same time.

In the rest of this paper, \sect{method} formulates the R-EBM for end-to-end ASR. Then \sectdot{setup} describes the data and models used for the experiments. Experimental results are presented in \sectdot{exp}. Further analysis on the behaviour of R-EBMs is given in \sectdot{analysis}, followed by conclusions in \sectdot{conclusion}.
\section{Residual Energy-Based Models}
\label{sec:method}
\subsection{Overview}
ASR models the conditional distribution of the text sequence $\vect{y}$ given the input acoustic sequence $\vect{X}$. For two state-of-the-art types of end-to-end trainable systems, namely attention-based sequence-to-sequence models and recurrent neural network transducers (RNN-Ts), the model distribution can be expanded using chain rule as in \eqndot{e2e}.
\begin{equation}
    P(\vect{y}|\vect{X}) = P(y_1|\vect{X})\prod_{t=2}^{T}P(y_t|y_{1:t-1},\vect{X}).
    \label{eqn:e2e}
    \vspace{-0.5em}
\end{equation}

These end-to-end models are locally normalised as an auto-regressive decoder predicts the next token based on the acoustic features and the past tokens for each step. The tokens used for end-to-end ASR models can be a set of characters, word pieces or even words. Given an unlimited model capacity, the auto-regressive model trained using maximum likelihood has the potential to learn the true data distribution perfectly~\cite{Deng2020Residual}. However, in practice, there are various drawbacks associated with locally normalised models. Specifically for end-to-end ASR, exposure bias~\cite{Ranzato2016SequenceLT} can be an issue during decoding when errors exist in the history; more search errors are likely to occur since only top few hypotheses are kept; and token-level maximum likelihood training does not directly minimise word error rates~\cite{Prabhavalkar2018MinimumWE}. Therefore, the distribution learned by a locally normalised auto-regressive model $P(\vect{y}|\vect{X})$ may not match the real data distribution $P_{\text{data}}$. To have a globally normalised model, \cite{Deng2020Residual} proposed residual energy-based models (R-EBM) that learns ``residual'' errors of the auto-regressive model to better match the data distribution. The R-EBM, parameterised by $\vect{\theta}$, can be formulated as
\begin{equation}
    P_{\vect{\theta}}(\vect{y}|\vect{X}) = \dfrac{P(\vect{y}|\vect{X})\exp\big(-E_{\vect{\theta}}(\vect{X}, \vect{y})\big)}{Z_{\vect{\theta}}(\vect{X})},
\end{equation}
where $P_{\vect{\theta}}$ is the joint model, $E_{\vect{\theta}}$ is the residual energy function, and $Z_{\vect{\theta}}$ is the partition function for the energy-based model, which can be computed as
\begin{equation}
    Z_{\vect{\theta}}(\vect{X}) = \sum_{y}P(\vect{y}|\vect{X})\exp\big(-E_{\vect{\theta}}(\vect{X}, \vect{y})\big).
    \vspace{-0.5em}
\end{equation}

\begin{figure}[t]
    \centering
    \includegraphics[width=0.85\linewidth]{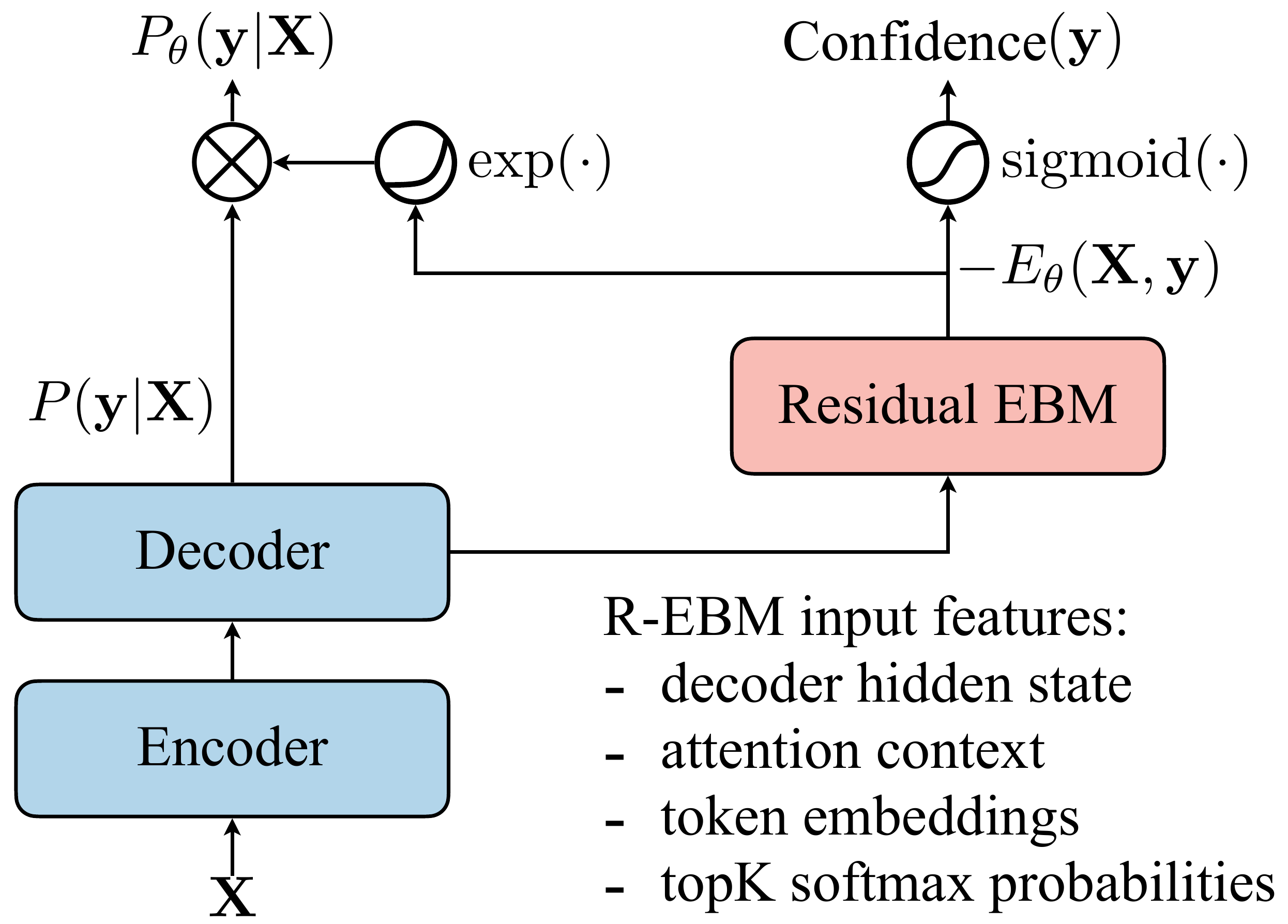}
    \caption{Schematic of a residual energy-based model (R-EBM) for an end-to-end ASR model. The baseline auto-regressive model is fixed during R-EBM training.}
    \label{fig:rebm}
    \vspace{-2em}
\end{figure}

In principle, the R-EBM itself can be any model that takes a pair of acoustic sequence $\vect{X}$ and hypothesis sequence $\vect{y}$ to produce a scalar value ($-E_{\vect{\theta}}(\vect{X},\vect{y})$). In this paper, the R-EBM is a sequence model that takes features including the current decoder hidden state, the acoustic context vector from the attention mechanism, the output token embeddings, and the topK softmax probabilities for each output step~\cite{li2020confidence}. Then the hidden representations for each output token are gathered by mean pooling before being passed to the output layer that uses sigmoid activation function. Since R-EBMs operate at the sequence-level, bi-directional models can be used.

\subsection{Training}
However, for a system with a token vocabulary size $V$ and a maximum output sequence length $T$, the partition function quickly becomes intractable as the summation is over $V^T$ possible sequences. With the baseline auto-regressive model $P$ fixed, conditional noise contrastive estimation (NCE)~\cite{Ma2018NoiseCE} can be used to train the R-EBM where the noise distribution is the auto-regressive ASR model $P$. The loss can be expressed as
\begin{align}
    \begin{split}
    \mathcal{L}(\vect{\theta}) = \mathbb{E}_{\vect{X}}\bigg\{&\mathbb{E}_{\vect{y}^{+}\sim P_{\text{data}(\cdot|\vect{X})}}\log\dfrac{1}{1+\exp\big(E_{\vect{\theta}}(\vect{X}, \vect{y}^{+})\big)}\\
    & + \mathbb{E}_{\vect{y}^{-}\sim P(\cdot|\vect{X})}\log\dfrac{1}{1+\exp\big(-E_{\vect{\theta}}(\vect{X}, \vect{y}^{-})\big)}\bigg\},
    \end{split}
\end{align}
where $\vect{y}^+$ are samples from the data distribution and $\vect{y}^-$ are samples from the noise distribution. For ASR, the noise samples are the $n$-best hypotheses from the auto-regressive model via beam search, and the positive samples are the corresponding ground truth transcription. Thus, the R-EBM is effectively a discriminator between incorrect sequences $Y^-$ and correct sequences $Y^+$, trained using the binary cross-entropy loss,
\begin{align}
    \begin{split}
        \mathcal{L}(\vect{X};\vect{\theta}) \approx & \,\,\dfrac{1}{|Y^+|}\sum_{\vect{y}\in Y^+}\log\dfrac{1}{1+\exp\big(E_{\vect{\theta}}(\vect{X}, \vect{y})\big)}\\
        & + \dfrac{1}{|Y^-|}\sum_{\vect{y}\in Y^-}\log\dfrac{1}{1+\exp\big(-E_{\vect{\theta}}(\vect{X}, \vect{y})\big)},
    \end{split}
\end{align}
where $Y^+\cup Y^- = \{\vect{y}_{\text{ground truth}}, \textsc{BeamSearch}(\vect{X}, n)\}$. Note that there may be more than one element in $Y^+$ since multiple tokenization using sub-word units for the ground truth may exist or the ground truth is among the $n$-best hypotheses. Finally, the parameters for the R-EBM can be estimated by optimising a binary classifier over all the utterances in the entire dataset.

\subsection{Inference}
For a given utterance during inference, the log-likelihood of a hypothesis and the negative energy value of the hypothesis are added to obtain the joint score as in \eqndot{rerank}.
\begin{equation}
    \log P_{\vect{\theta}}(\vect{y}|\vect{X})\propto\log P(\vect{y}|\vect{X}) - E_{\vect{\theta}}(\vect{X},\vect{y}).
    \label{eqn:rerank}
\end{equation}
With a shared partition function, the $n$-best hypotheses can be re-ranked based on the joint scores to yield the best candidate.

From another perspective, the R-EBM is a binary classifier that learns to assign scores close to 1 for correct hypotheses and 0 for erroneous ones, which is also the objective for utterance-level confidence scores~\cite{Kumar2020UtteranceCM}. Confidence scores can be used to automatically assess the quality of transcriptions of ASR systems. For applications where utterance-level confidence scores are also required, the R-EBM can be used to achieve two aims. The pre-sigmoid values of R-EBM $(-E_{\vect{\theta}}(\vect{X},\vect{y}))$ can be used to rerank $n$-best hypotheses to lower the word error rate of the ASR system while the post-sigmoid values can be used as a model-based confidence measure.

\section{Experimental Setup}
\label{sec:setup}
\subsection{Data}
For training the baseline ASR model and R-EBMs, the ``train-clean-100'' subset of the LibriSpeech corpus~\cite{Panayotov2015LibrispeechAA} was used. The 100-hour training set has 28.5k utterances and the average duration per utterance is 12.7 seconds. The dev and test sets are dev-clean/dev-other and test-clean/test-other with each one being over 5 hours. This dataset contains read speech from audiobooks. The text data for the language model (LM) is the LibriSpeech LM corpus with 40 million sentences. The acoustic features are 80-dimensional filterbank coefficients with $\Delta$ and $\Delta\Delta$ and the modelling units are a set of 1024 word-pieces~\cite{Schuster2012JapaneseAK} derived from the LibriSpeech 100h training transcriptions.

\subsection{Models}
The baseline model architecture is an attention-based sequence-to-sequence model, where the encoder has 2 convolutional layers with a stride of 2 followed by a 4-layer bi-directional long short-term memory (LSTM) network with 1024 units in each direction. The decoder has a 2-layer uni-directional LSTM network with 1024 units. The total number of parameters is 145 million. The Adam optimiser is used with a learning rate of 0.001 and batch size of 512. Training techniques such as SpecAugment~\cite{Park2019SpecAugmentAS}, dropout, label smoothing, Gaussian weight noise and exponential moving average are used to improve performance. The LM has 2 uni-directional LSTM layers with 1024 units in each layer. Shallow fusion~\cite{Glehre2015OnUM} is used for decoding and for generating $n$-best hypotheses for R-EBMs. Hyper-parameters for the attention-based models, the language model and beam search are tuned on the dev sets. 

The R-EBMs are trained with the baseline ASR model fixed. The $n$-best hypotheses of the training set are generated on-the-fly with beam size $n$ and random SpecAugment masks. For time masks, instead of 2 masks with a maximum of 40 frames per mask for the baseline ASR model, 10 masks with a maximum of 50 frames per mask are used for R-EBM training. This is to simulate the errors made by the model during inference and the randomness of masks allows diverse errors to appear during training. The WER on the augmented training set should ideally match that of the dev set.
\section{Experimental Results}
\label{sec:exp}
\subsection{Length Normalisation and Log-linear Interpolation}
\begin{table}[ht]
    \vspace{-1em}
    \centering
    \caption{Impact on WERs (\%) by applying length normalisation (LN) for log-likelihood score on dev-clean/dev-other sets. The R-EBM is a uni-directional LSTM and the beam size is 8.}
    \begin{tabular}{lccc}
        \toprule
         & 1-best & R-EBM & joint \\
        \midrule
        w/o LN & 5.84/18.48 & \multirow{2}{*}{5.44/18.13} & 5.38/17.91\\
        w/ LN & 5.28/18.41 &  & 5.09/17.77\\
        \bottomrule
    \end{tabular}
    \vspace{-1em}
    \label{tab:ln}
\end{table}
After beam search, $n$-best hypotheses are determined by keeping the $n$ terminated hypotheses with the highest sequence-level log-likelihood. However, this criterion may favour shorter hypotheses when finding the 1-best. Therefore, normalising the log-likelihood by the number of tokens in each hypothesis results in a slightly lower WER as shown in \tbl{ln}. Length normalisation (LN) becomes more important when the $n$ is large. When combining the log-likelihood score $\log P(\vect{y}|\vect{X})$ with the negative energy score $-E_{\vect{\theta}}(\mathbf{X},\mathbf{y})$ for the joint score, a coefficient $\alpha$ is tuned on dev sets to minimise WER and accommodate potentially different numerical ranges as in \eqn{ln}.
\begin{equation}
    \hat{\vect{y}} = \argmax_{\vect{y}}\,\,\dfrac{\log P(\vect{y}|\vect{X})}{|\vect{y}|} - \alpha E_{\vect{\theta}}(\vect{X},\vect{y}).
    \label{eqn:ln}
\end{equation}

\tbl{ln} shows that ranking the $n$-best hypotheses just using the R-EBM scores reduces the WER compared to 1-best WER without LN. After log-linear interpolation, the WER of the joint model is lower than the 1-best results with or without LN. Therefore, all the following experiments will use LN.

\subsection{R-EBM Architecture}
\begin{table}[ht]
    \vspace{-1em}
    \centering
    \caption{Comparison of WERs (\%) by using uni-directional and bi-directional LSTMs for R-EBMs on dev-clean/dev-other sets. The beam size for decoding is 8.}
    \begin{tabular}{lccc}
        \toprule
         & 1-best & R-EBM & joint \\
        \midrule
        uni-directional & \multirow{2}{*}{5.28/18.41} & 5.44/18.13 & 5.09/17.77\\
        bi-directional & & 5.33/18.05 & 5.07/17.69\\
        \bottomrule
    \end{tabular}
    \vspace{-1em}
    \label{tab:archi}
\end{table}
As a globally normalised model, an R-EBM can be bi-directional and take advantage of the full context in the hypotheses. \tbl{archi} compares the performance of uni-directional and bi-directional R-EBMs. Both R-EBMs have two layers of LSTMs with 512 units in each direction. Since the bi-directional model performs slightly better, all the following experiments will use bi-directional LSTMs for R-EBMs.

\subsection{Effect of the Size of N-best Lists}
\begin{table}[ht]
    \vspace{-1em}
    \centering
    \caption{WERs of the ASR model and joint models with various numbers of hypotheses used for training and inference.}
    \begin{tabular}{cr|rrrr|r}
        \toprule
        & $n$ & oracle & 1-best & R-EBM & joint & WERR \\
        \midrule
              & 4  &  4.19	&  5.41	&  5.58	&  5.30	& 2.0 \\
        dev-  & 8  &  3.52	&  5.28	&  5.33	&  5.07	& 4.0 \\
        clean & 16 &  3.03	&  5.17	&  5.23	&  4.86	& 6.0 \\
              & 32 &  2.68	&  5.23	&  5.33	&  4.80	& 8.2 \\
        \midrule
              & 4  & 17.10	& 19.04	& 19.07	& 18.71	& 1.7 \\
        dev-  & 8  & 15.33	& 18.41	& 18.05	& 17.69	& 3.9 \\
        other & 16 & 13.96	& 18.06	& 17.67	& 17.12	& 5.2 \\
              & 32 & 12.93	& 17.98	& 17.31	& 16.75	& 6.8 \\
        \bottomrule
    \end{tabular}
    \vspace{-1em}
    \label{tab:nbest}
\end{table}
The performance ceiling of the joint model is the oracle WER of the $n$-best hypotheses. In this section, the lengths of $n$-best lists range from 4 to 32 for both training and inference. \tbl{nbest} shows that oracle WERs improve consistently when larger $n$-best lists are used whereas 1-best WERs only has minor reduction. With more hypotheses available, the joint WERs reduces significantly together with WERs ranked by R-EBM scores only. The last column in \tbl{nbest} shows that WER relative reduction (WERR) of the joint model over 1-best steadily increases with larger $n$-best lists, which indicates that the gain from the R-EBMs outpaces that of the 1-best. With 32-best, 8.2\%/6.8\% WERRs are obtained for dev-clean/dev-other sets. Top 32-best will be used for all the following experiments.

\subsection{Recognition and Confidence Estimation Performance}
\begin{table}[ht]
    \vspace{-1em}
    \centering
    \caption{Recognition and confidence performance on LibriSpeech test sets. The average of token-level scores from CEM is used as utterance-level scores. For both CEM and R-EBM, the best interpolation coefficients are tuned on the dev sets.}
    \begin{tabular}{lccccc}
        \toprule
         & \multicolumn{2}{c}{test-clean} && \multicolumn{2}{c}{test-other} \\
         \cmidrule{2-3}\cmidrule{5-6}
         & WER $\downarrow$ & AUC $\uparrow$ && WER $\downarrow$ & AUC $\uparrow$\\
        \midrule
        baseline  & 5.61 & 0.684 && 18.68 & 0.529\\
        \;+ CEM~\cite{li2020confidence}   & 5.59 & 0.697 && 18.44 & 0.501\\
        \;+ R-EBM & \bf{5.15} & \bf{0.770} && \bf{17.42} & \bf{0.679}\\
        \bottomrule
    \end{tabular}
    \vspace{-1em}
    \label{tab:results}
\end{table}
Based on previous experimental results on dev sets, bi-directional LSTMs with 32-best hypotheses and length normalisation are used as the best setup. By applying the log-linear combination coefficients $\alpha$ tuned on dev-clean/dev-other sets, the results on two test sets are shown in \tbl{results}. The area under the precision-recall curve (AUC) is used as the metric for confidence estimation~\cite{Ragni2018ConfidenceEA,li2020confidence}. AUC is a number between 0 and 1 and higher values indicate more reliable confidence scores. Also in \tbl{results}, the second row corresponds to the confidence estimation module (CEM)~\cite{li2020confidence}, which predicts a confidence score for each token in the hypothesis sequence using the same input features as R-EBMs. By averaging the token-level scores\footnote{For each hypothesis, the mean of the pre-sigmoid logits of all tokens is used for $n$-best reranking, whereas the mean of the post-sigmoid confidence scores of all tokens is used for confidence evaluation.}, utterance-level scores can be used to combine with the baseline model. Note that although the CEM yields improved confidence at token and word levels as in \cite{li2020confidence}, the utterance-level confidence may under-perform the baseline log-likelihood score. Since the R-EBM is directly optimised for the utterance-level confidence, issues such as multiple tokenisations for the same word or sequence and deletion errors are addressed implicitly during training. As a result, the R-EBM reduces WERs and significantly improves utterance-level confidence. 

\subsection{Scalability}
\label{sec:w2v}
\begin{table}[ht]
    \vspace{-1em} 
    \centering
    \caption{Recognition and confidence performance on LibriSpeech test sets when the encoder is initialised using a pre-trained wav2vec 2.0 (w2v2) as a stronger baseline.}
    \begin{tabular}{lccccc}
        \toprule
         & \multicolumn{2}{c}{test-clean} && \multicolumn{2}{c}{test-other} \\
         \cmidrule{2-3}\cmidrule{5-6}
         & WER $\downarrow$ & AUC $\uparrow$ && WER $\downarrow$ & AUC $\uparrow$\\
        \midrule
        w2v2                        & 2.63  & 0.786 && 4.74 & 0.684\\
        \;+ R-EBM                          & \bf{2.49} & \bf{0.928} &&  \bf{4.53} & \bf{0.890}\\
        \bottomrule
    \end{tabular}
    \vspace{-1em}
    \label{tab:wav2vec}
\end{table}
This section investigates the situation when the auto-regressive ASR model has seen much more data such that the baseline WER is far lower. In this set of experiments, the encoder of the ASR model is first initialised with the pre-trained wav2vec 2.0 (w2v2) model~\cite{Baevski2020wav2vec2A} trained on 57.7 thousand hours of unlabelled speech data from Libri-light~\cite{Kahn2020LibriLightAB} and then fine-tuned on the same amount of labelled data (``train-clean-100'') as before\footnote{Out implementation follows \cite{Zhang2020PushingTL}, which shows the state-of-the-art performance on LibriSpeech.}. Although the WERs in \tbl{wav2vec} are much lower than in \tbl{results}, the joint model further yields 5.3\%/4.4\% WERRs on test-clean/test-other. Meanwhile, AUCs are much higher with more unlabelled data as expected, but R-EBMs can furthur boost confidence estimation significantly.

\section{Analysis}
\label{sec:analysis}
\subsection{Relative Improvement by Utterance Length}
\begin{figure}[ht]
    \vspace{-1em}
    \centering
    \begin{subfigure}{.49\linewidth}
        \centering
        \includegraphics[width=\textwidth]{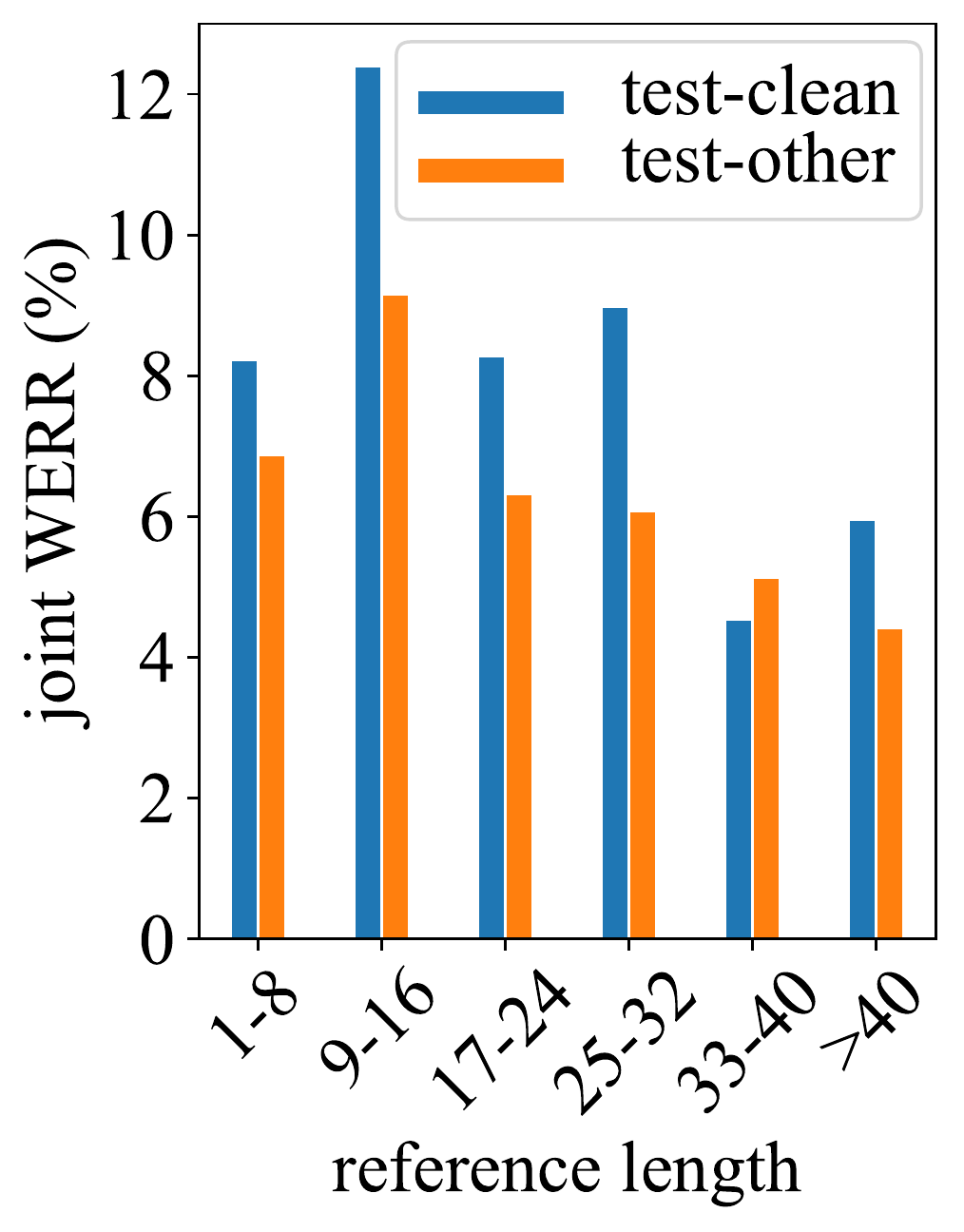}
        \label{fig:werr_joint}
    \end{subfigure}%
    \begin{subfigure}{.49\linewidth}
        \centering
        \includegraphics[width=\textwidth]{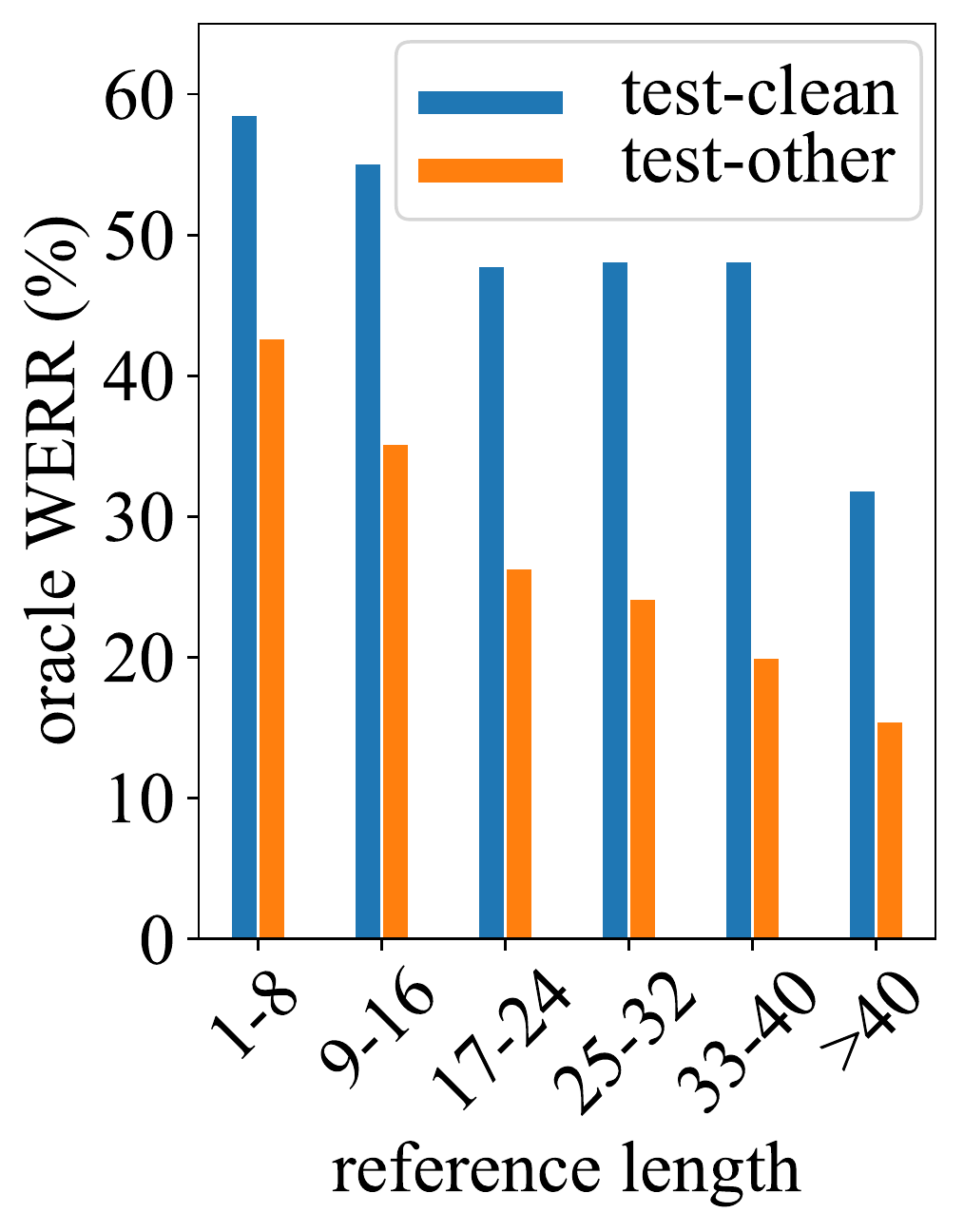}
        \label{fig:werr_oracle}
    \end{subfigure}
    \vspace{-2em}
    \caption{WERR with respect to the number of words in reference sequences. Left is the WERR of joint over 1-best and right is the WERR of oracle over 1-best. 32-best hypotheses are used.}
    \vspace{-1em}
    \label{fig:werr}
\end{figure}
\fig{werr} shows the breakdown of WERR for the joint model and the oracle hypotheses with respect to the number of words in reference sequences. Oracle WERR is lower for longer utterances as the number of alternatives per word is fewer with a given number of top hypotheses. The general trend of the joint WERR follows the trend of the oracle WERR except for short utterances (1-8 words in reference). We hypothesise that R-EBMs may need more global context information to give a higher WER reduction.

\subsection{Distribution Matching}
\begin{figure}[ht]
    \vspace{-1em}
    \centering
    \begin{subfigure}{.49\linewidth}
        \centering
        \includegraphics[width=\textwidth]{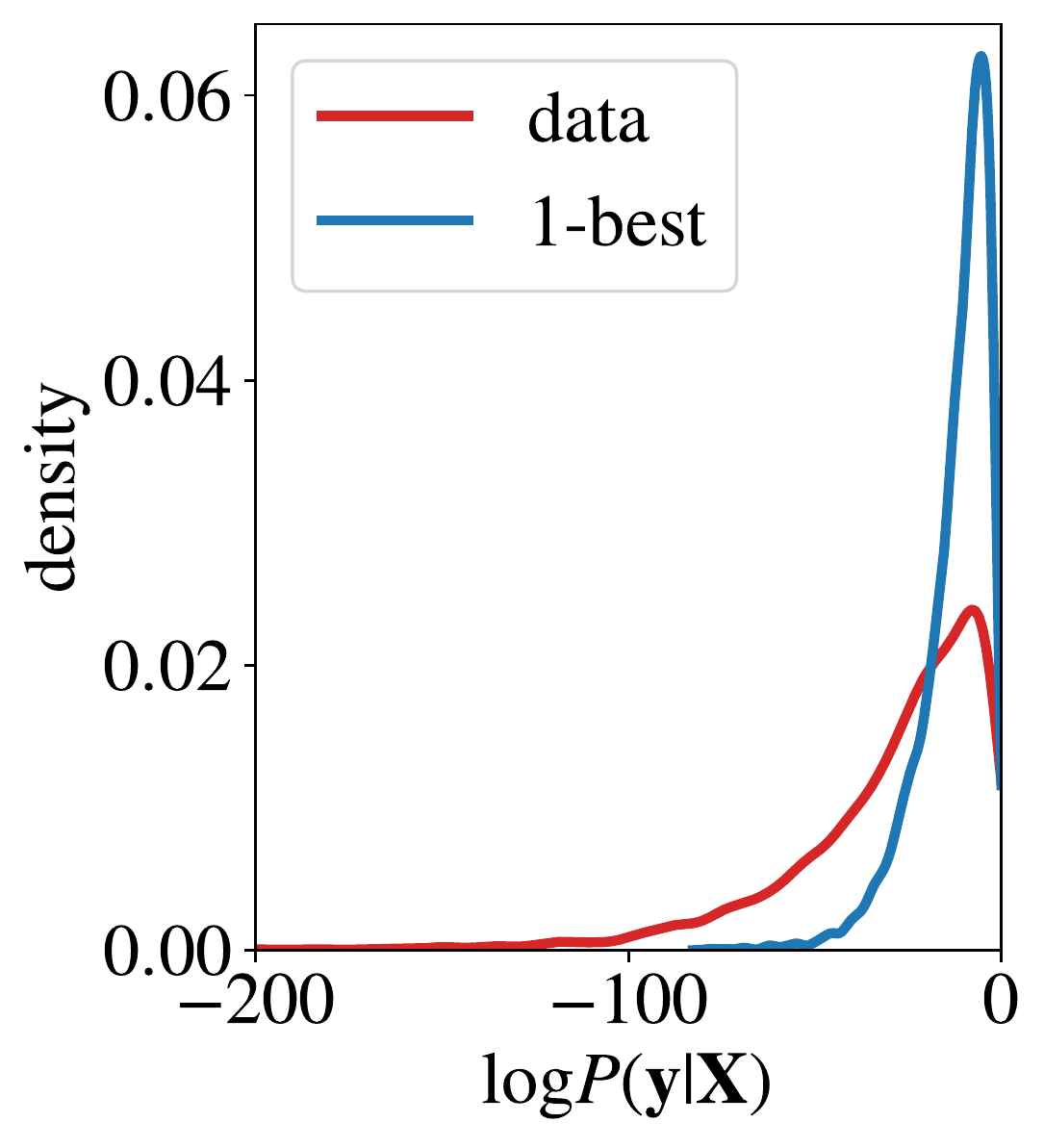}
        \label{fig:test_other_ll}
    \end{subfigure}%
    \begin{subfigure}{.485\linewidth}
        \centering
        \includegraphics[width=\textwidth]{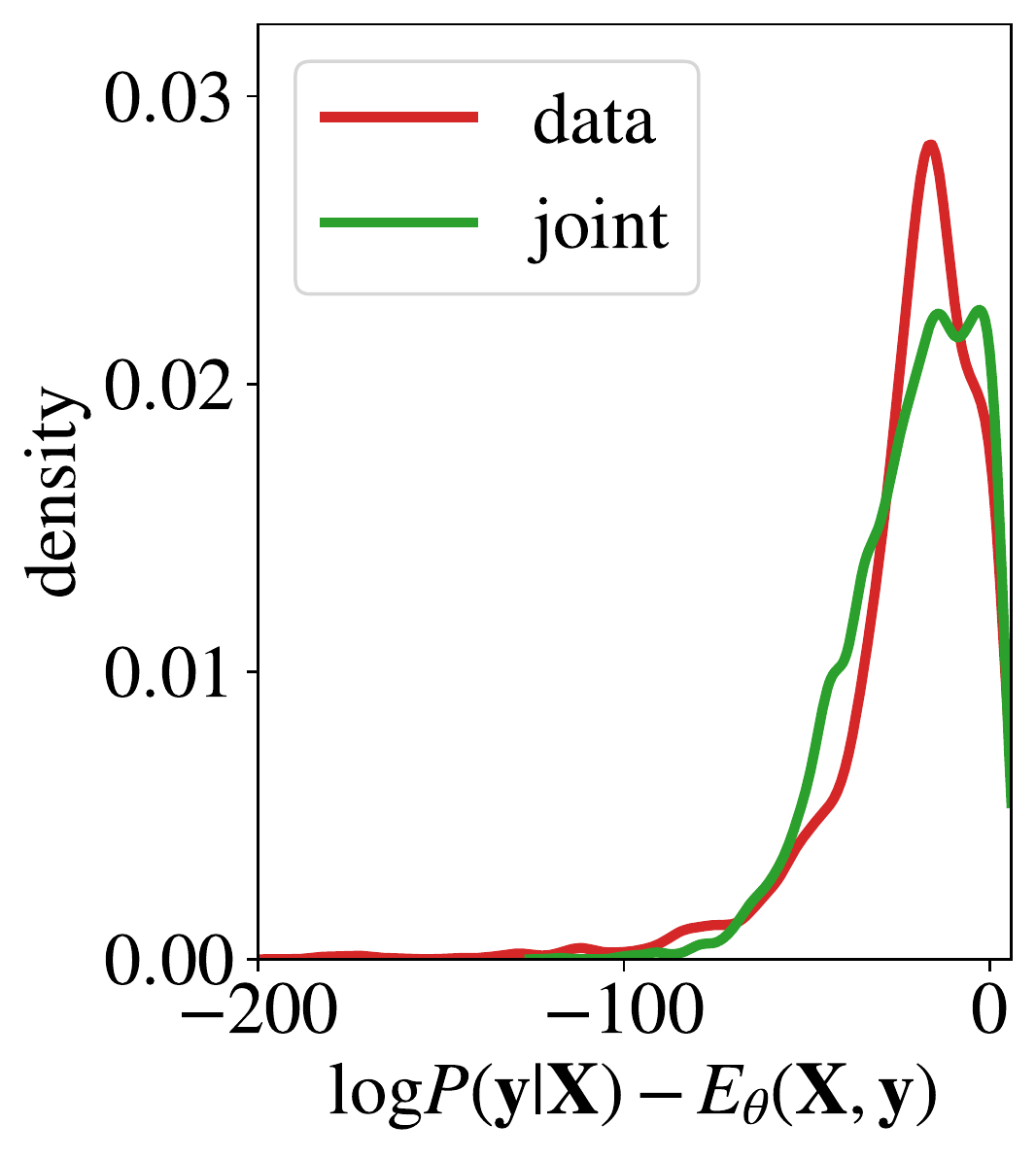}
        \label{fig:test_other_joint}
    \end{subfigure}
    \vspace{-2em}
    \caption{Density plot of log-probability scores using the baseline model (left) and the joint model (right) on test-other set.}
    \vspace{-1em}
    \label{fig:distribution}
\end{figure}
If the joint model $P_{\vect{\theta}}$ matches the data distribution $P_{\text{data}}$ better, then statistics computed on a large set of samples from the two distributions should also match~\cite{Baevski2020wav2vec2A}. \fig{distribution} shows the density plot of the log-likelihood scores (left) and the joint model scores (right) on test-other set. Red lines correspond to the score distributions of the ground truth transcriptions. The distribution of log-likelihood scores of the best hypotheses from the auto-regressive model does not match the data distribution well. However, the distribution from the joint model is much closer to the data distribution.
\section{Conclusions}
\label{sec:conclusion}
This paper proposes to use residual energy-based models (R-EBMs) to complement locally normalised auto-regressive end-to-end ASR models. R-EBM is globally normalised as it learns from the residual error of the locally normalised model. R-EBMs can also be viewed as an utterance-level confidence estimator for ASR. Experiments show that R-EBMs can reduce speech recognition error rates while improving the confidence scores at the utterance level, even on top of a state-of-the-art baseline model trained using wav2vec 2.0. Further analysis shows that the performance of R-EBM may depend on the amount of context, and confirms that the R-EBM closes the gap between the model distribution and the data distribution.

\clearpage
\bibliographystyle{IEEEtran}
\bibliography{refs}

\end{document}